\makeatletter
\declare@file@substitution{revtex4-1.cls}{revtex4-2.cls}
\makeatother

\documentclass[preprint]{aastex631}

\usepackage{natbib}
\usepackage{url}
\usepackage{xcolor}  
\usepackage{lineno}

\begin{document}

\title{TIC 322208686: An Eclipsing System with $\gamma$ Doradus Pulsations and a Third Component on a Wider Orbit }
\correspondingauthor{Jae Woo Lee}
\email{jwlee@kasi.re.kr}
\author[0000-0002-5739-9804]{Jae Woo Lee}
\affil{Korea Astronomy and Space Science Institute, Daejeon 34055, Republic of Korea}

\author[0000-0002-8692-2588]{Kyeongsoo Hong}
\affil{Korea Astronomy and Space Science Institute, Daejeon 34055, Republic of Korea}

\author[0000-0002-8394-7237]{Min-Ji Jeong}
\affil{Korea Astronomy and Space Science Institute, Daejeon 34055, Republic of Korea}

\author[0000-0001-9339-4456]{Jang-Ho Park}
\affil{Korea Astronomy and Space Science Institute, Daejeon 34055, Republic of Korea}

\author[0000-0003-1916-9976]{Pakakaew Rittipruk}
\affil{National Astronomical Research Institute of Thailand, Chiang Mai 50200, Thailand}

\author[0000-0002-6687-6318]{Hye-Young Kim}
\affil{Korea Astronomy and Space Science Institute, Daejeon 34055, Republic of Korea}
\affil{Department of Astronomy and Space Science, Chungbuk National University, Cheongju 28644, Republic of Korea}

\begin{abstract}
TIC 322208686 is known to be a detached binary that exhibits two types of variability: pulsation and eclipse. We present 
the physical properties of the target star using the short-cadence TESS data from sectors 24, 57, and 58, and our echelle 
spectra that show the presence of a tertiary companion. The spectral analysis led to the triple-lined radial velocities and 
the atmospheric parameters of the eclipsing components. Joint modeling of these observations reveals that the eclipsing pair 
contains two F-type stars with masses $1.564\pm0.012$ $M_\odot$ and $1.483\pm0.012$ $M_\odot$, 
radii $1.588\pm0.011$ $R_\odot$ and $1.500\pm0.012$ $R_\odot$, effective temperatures $7028\pm100$ K and $7020\pm110$ K, and 
luminosities $5.51\pm0.32$ $L_\odot$ and $4.90\pm0.32$ $L_\odot$. The light contributions of the three stars obtained from 
this modeling match well with those calculated from the observed spectra. The binary star parameters are 
in satisfactory agreement with evolutionary model predictions for age $t$ = 0.4 Gyr and metallicity $Z$ = 0.03. We extracted 
11 significant frequencies from the TESS light residuals with the binary effects removed. Of these, five signals between 
0.65 day$^{-1}$ and 1.89 day$^{-1}$ can be considered as $\gamma$ Dor pulsations originating mainly from the primary component, 
while the other frequencies are likely instrumental artifacts or combination terms. These results suggest that TIC 322208686 
is a hierarchical triple, containing a pulsating eclipsing pair and a tertiary companion. 
\end{abstract}

\keywords{Eclipsing binary stars (444) --- Spectroscopic binary stars (1557) --- Multiple stars (1081) --- Fundamental parameters of stars (555) 
--- Stellar pulsations (1625) --- Gamma Doradus variable stars (2101) --- Stellar evolution (1599)}

\section{Introduction} \label{sec_intro}

Our knowledge of stellar evolution requires precise measurements of absolute properties such as mass, radius, and effective 
temperature, which are invaluable in many fields of astrophysics. These measurements are cornerstones for calibrating theoretical 
evolution models and advancing our understanding of stellar physics. Eclipsing binaries (EBs) serve as the primary source for 
deriving the physical parameters of their component stars in various stages of evolution. Moreover, they have potential as 
distance indicators for nearby galaxies of which they are members \citep{Hilditch2001,Pietrzynski+2013}. 
Orbital solutions that employ different types of observations have significant advantages over those using only one type. 
For example, combining photometry and spectroscopy allows for accurate determination of fundamental parameters without 
any assumptions. The absolute masses and radii of detached EBs can be determined with better than 3 \% precision through 
joint modeling of their double-lined radial velocities (RVs) and light curves \citep{Andersen1991,Torres+2010}. Such EBs are 
representative of normal stars, as they have not undergone mass transfer. Modern space-based photometric and echelle spectroscopic 
studies allow measurements of mass and radius of each component with precisions of $\sim$0.2 \% \citep[e.g.,][]{Maxted+2020}. 

The study of stellar pulsations, known as asteroseismology, can also help improve theoretical models of stars. Comparing 
the observed pulsation frequencies to theoretical predictions can provide information about the interior physics of stars 
from core to surface, including their ages, densities, rotational profiles, and chemical compositions, etc \citep{Aerts+2010}. 
Therefore, EBs that exhibit oscillating signals are highly valuable due to the synergistic effect of pulsation and binary 
properties \citep{Murphy2018,Kurtz2022}. Some detached EBs exhibit pulsating features in one or both components 
\citep[e.g.,][]{Southworth+2022b}. Recently, many new pulsating EBs have been discovered through space-based photometric archives 
such as Kepler and TESS \citep{Gaulme+2019,Chen+2022,Kahraman+2022,Shi+2022,Eze+2024}. However, most of their physical properties 
remain poorly defined due to lack or absence of spectroscopic data. 

This study focuses on the TESS target TIC 322208686 (BD+67 1348, TYC 4462-599-1, 2MASS J21413664+6745394, WISE J2141366+674539, 
Gaia DR3 2222840880731556736; $T_{\rm p}$ = $+$9.303), a detached EB with an eclipse period of 1.354$\pm$0.005 days 
discovered by \citet{Chen+2018} in the WISE data. \citet{Shi+2022} detected a dominant pulsation with a cycle of 0.79216$\pm$0.00004 days 
and an amplitude of 10.11$\pm$0.03 mmag in the EB system, based on 120-s sampling data collected from TESS Sector 24. They reported 
that the pulsating signal does not follow the relation between the orbital ($P_{\rm orb}$) and pulsation ($P_{\rm pul}$) periods 
of oscillating eclipsing Algols \citep{Mkrtichian+2004,Soydugan+2006,Liakos+2012,Zhang+2013}. Our primary goal here is to 
accurately determine the absolute properties of the triple-lined EB system with a tertiary companion and multiperiodic pulsations, 
and to investigate its evolution.

\section{TESS Photometry and Eclipse Timings} \label{sec_photometry}

\begin{figure}
\includegraphics[scale=0.9]{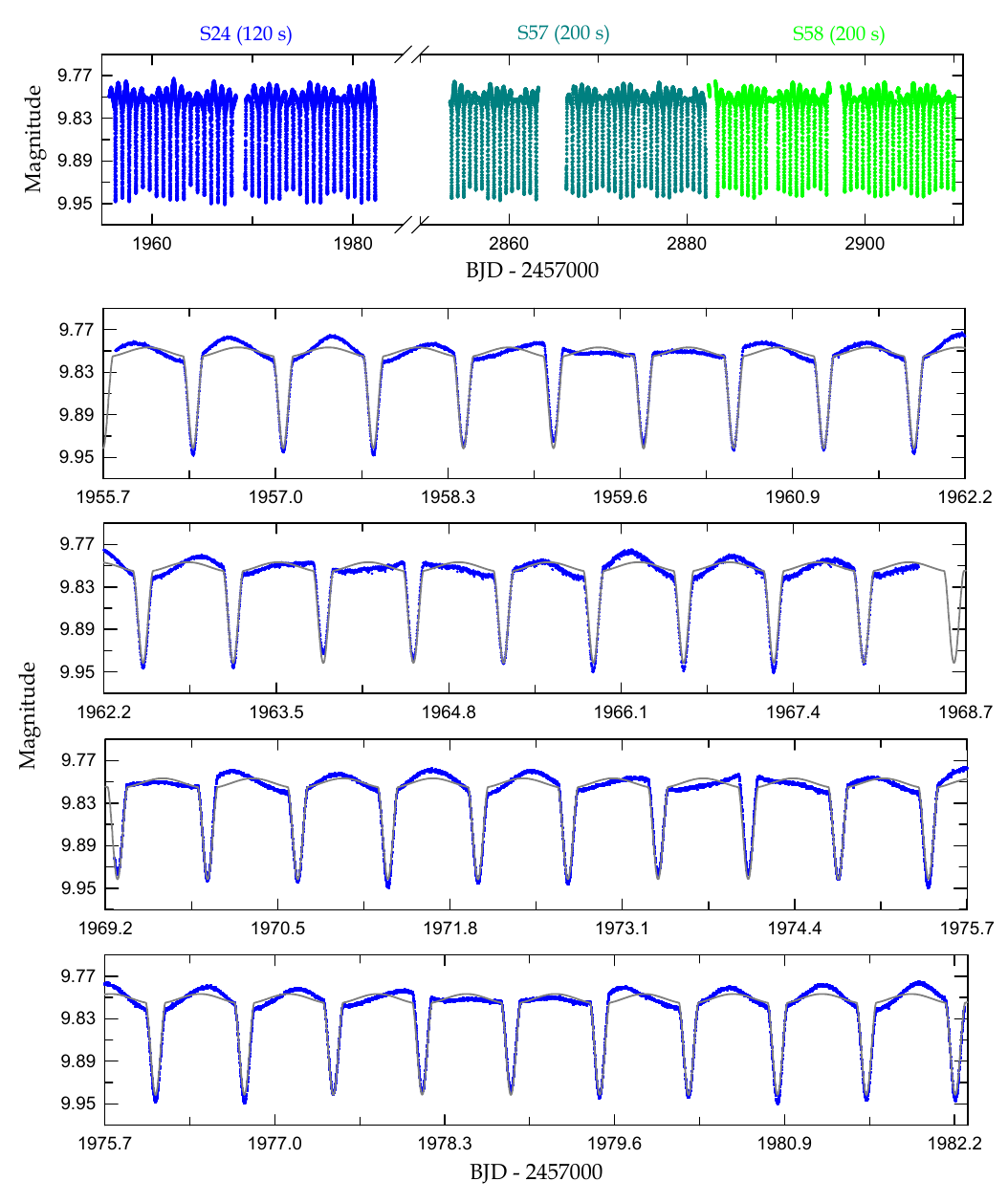}
\caption{Time-series TESS data of TIC 322208686 observed during Sectors 24 (blue), 57 (cyan), and 58 (green). 
The top panel shows all observations from the three sectors, and the other panels detail 120-s cadence measurements 
from Sector 24. The gray lines represent the synthetic model curves computed from our binary star parameters in Table \ref{Tab3}. }
\label{Fig1}
\end{figure}

TIC 322208686 was monitored by the TESS satellite \citep{Ricker+2015} and recorded with 30-min cadence in Sectors 16, 17, and 
18 (S16$-$18; 2019 September 12 to November 27), 120-s cadence in Sector 24 (S24; 2020 April 16 to May 12), and 200-s cadence 
in Sectors 57 and 58 (S57$-$58; 2022 September 30 to November 26). In this work, we concentrated on the short-cadence SAP 
observations of S24 and S57$-$58 retrieved from the MAST archive\footnote{\url{https://archive.stsci.edu/}. The TESS data used 
are accessible at \url{http://dx.doi.org/10.17909/kzxc-yy89}.} for binary modeling and pulsation analysis. 
Their CROWDSAP parameter is 0.9865$\pm$0.0066 on average. The detrending and flux-to-magnitude conversion of the TESS data 
are identical to those described and applied in \citet{Lee+2024}. The resulting time-series observations of TIC 322208686 are 
displayed in Figure \ref{Fig1}, where the observation gap between S24 and S57 is about 871 days. As reported in \citet{Shi+2022}, 
it is clear that our program target exhibits pulsating signals, especially in outside eclipse phases. We finally secured 18,216 
and 21,631 observations in 120-s and 200-s sampling modes, respectively, with a total time span of 954 days. 

\begin{figure}
\includegraphics[scale=1.0]{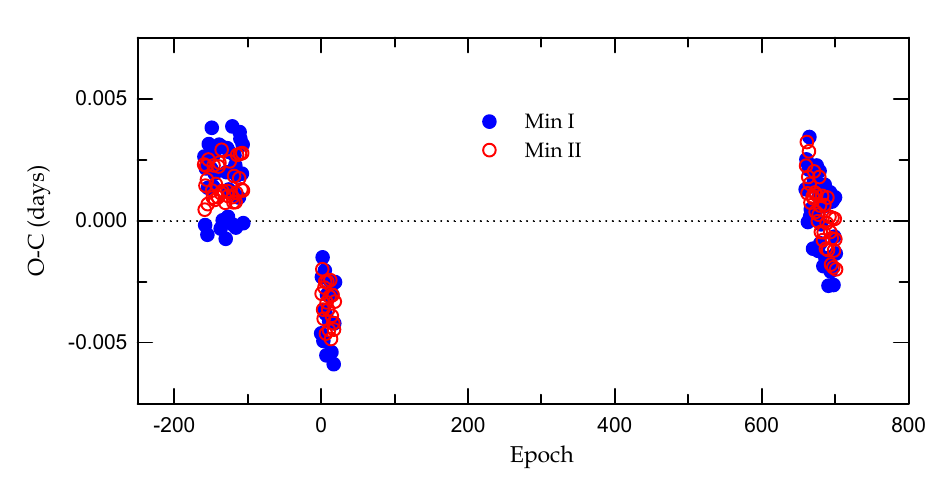}
\caption{Eclipse timing $O$--$C$ residuals for TIC 322208686 constructed with the orbital ephemeris (1). The filled and 
open circles represent the times of the primary and secondary eclipses, labeled Min I and Min II, respectively. }
\label{Fig2}
\end{figure}

We measured 92 primary (Min I) and 88 secondary (Min II) minimum epochs of TIC 322208686 from each eclipse curve 
of the available TESS data, including those with 30-min cadence \citep{Kwee+1956}. These calculations used at least 8 consecutive 
observations for each 30-min minimum, and about 100 and 60 observations for 120-s and 200-s minima, respectively. 
Due to the limited data points, the errors for the 30-min measurements are likely underestimated. In order to obtain the binary's 
orbital ephemeris, we applied linear least-squares fitting to them, which resulted in the following ephemeris:
\begin{equation}
\mbox{Min I} = \mbox{BJD}~ 2,458,956.38202(20) + 1.35931375(46)E,
\end{equation}
where the parenthesized numbers are the 1-$\sigma$ errors of the last two digits. 
The orbital period is slightly shorter and more precisely derived than that (1.35964$\pm$0.00032) of \citet{Shi+2022}. 
The observed TESS eclipse timings are compiled in Table \ref{Tab1}, where the third and fourth columns are the cycle numbers 
and $O-C$ values calculated with equation (1). The $O-C$ diagram of Figure \ref{Fig2} shows that although the mid-eclipse timings 
of TIC 322208686 are somewhat scattered by its pulsating features, there is a noticeable change in the orbital period. 
The timing variation may be due to the third object detected spectroscopically in this study. Additionally, the primary and 
secondary minima are in phase with each other, suggesting that the binary may be in a circular orbit. 


\section{BOES Spectroscopy and Data Analysis} \label{sec_spectroscopy}

Time-series echelle spectroscopy of TIC 322208686 was performed on 2023 November 12 and 14 at BOAO, Korea. The observations 
were made using the 1.8-m telescope and the BOES spectrograph \citep{Kim+2007}, providing a spectral range from 3600 to 
10,200 \AA~ and the resolving power of $\sim$30,000. A total of 16 optical spectra were acquired with an integration time 
of 1800 s per frame, which is equivalent to 0.015 of the orbital period, and is therefore unaffected by orbital smearing. 
In addition, images for pre-processing (bias and Tungsten-Halogen lamp) and wavelength calibration (Th-Ar lamp) were 
collected before and after these target spectra. The observed BOES spectra were reduced with the \texttt{echelle} package 
of the \texttt{IRAF} software with the standard procedures such as spectral extraction, wavelength calibration, and 
normalization by continuum. Their signal-to-noise ratios (SNR) were found to be about 30 at a wavelength of around 5000 \AA. 

\begin{figure}
\includegraphics[scale=0.75]{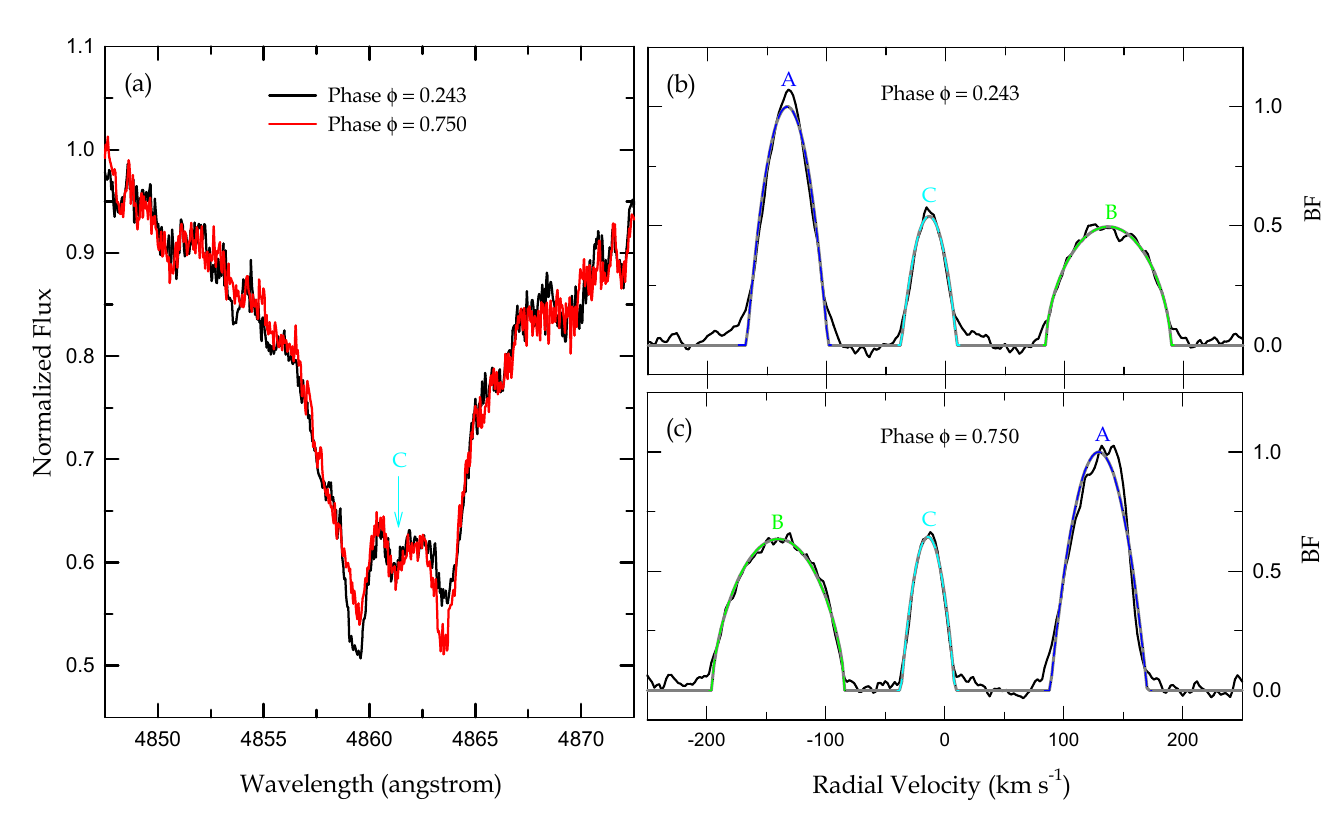}
\caption{Left panel (a): H$_{\rm \beta}$-region spectra observed at orbital phases $\phi$ = 0.243 (BJD 2,460,263.0198) and 
$\phi$ = 0.750 (BJD 2,460,260.9902). The signature of the tertiary companion, indicated by an arrow, is well defined between 
the two absorptions of the eclipsing pair AB. Right panels (b) and (c): BF profiles for the same phases as before obtained 
using the \texttt{RaveSpan} code. The solid black lines indicate the BF profiles, which exhibit triple peaks and are fitted 
with three rotational profile functions representing the primary (blue), secondary (green), and third (cyan) components. 
The gray lines correspond to the combination of these rotational profiles. }
\label{Fig3}
\end{figure}

The observed spectra of TIC 322208686 clearly show the three absorption lines originating from the eclipsing components and 
an additional object. The sample spectra in the H$_{\rm \beta}$ region, presented in the left panel of Figure \ref{Fig3}, demonstrate 
that the absorption lines of the close pair AB are shifted due to the Doppler effect, while those of the third spectroscopic source 
are unchanged. For the RV measurements of the three components, we adopted the broadening function (BF) technique \citep{Rucinski2002} 
of the \texttt{RaveSpan} software\footnote{\url{https://users.camk.edu.pl/pilecki/ravespan/}} \citep{Pilecki+2012}, which is 
especially useful when the spectral lines are blended. The BF method decomposes a target composite spectrum into individual stars 
using a sharp-lined template, maintains higher resolution compared to the cross-correlation technique, and makes it easier to 
determine the light contribution by simple integration. In this way, it enables the identification of faint companions \citep[e.g.,][]{Lee+2024}. 

\begin{figure}
\includegraphics[]{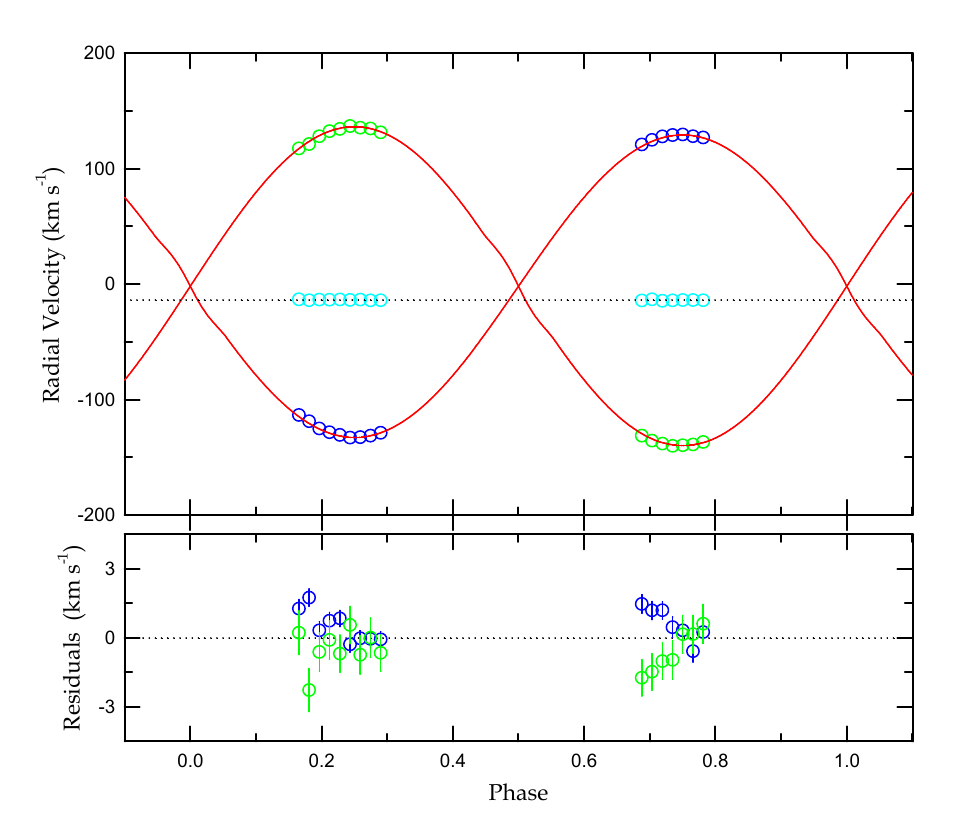}
\caption{RV curves of TIC 322208686 with fitted models. The blue, green, and cyan circles are the RV measurements for the primary 
(A), secondary (B), and third (C) components, respectively. The red solid curves represent the results of the W-D binary model, 
and the dotted line denotes the average RV of $-$13.86 km s$^{-1}$ for the third component. The lower panel shows the residuals 
between observations and models. 
} 
\label{Fig4}
\end{figure}

In this process, the template spectrum was taken from the model library of synthetic spectra by \citet{Coelho+2005}. 
The spectral region from 4840 to 5385 \AA~ containing H$_{\rm \beta}$ and Mg I b triplet at 5167, 5172, and 5183 \AA~ was 
selected for the BF work, where no significant telluric lines were observed. We measured the RVs applying the rotational function 
to each BF profile. As samples, the BF profiles for two orbital phases are illustrated with the fitted rotational profiles in 
the right panels of Figure \ref{Fig3}. The RV measurement results for this triple are summarized in Table \ref{Tab2} and 
displayed in Figure \ref{Fig4}. In this figure, the RVs of the eclipsing pair vary with its orbital motion, while those of 
the tertiary companion remain almost constant at the average RV of $-13.86\pm0.37$ km s$^{-1}$. 
The light contributions of individual components, $l/(l_{\rm A}+l_{\rm B}+l_{\rm C})$, estimated from 
the BF profiles are on average 44 \%, 40 \%, and 16 \% for the primary, secondary, and tertiary stars, respectively 
\citep{Rucinski+2008,Lee+2024}. These values are intensity ratios of the spectral lines used in the BF and therefore do not 
directly constrain the continuum or overall light contributions. 

To measure the components' atmospheric parameters, the observed spectra were disentangled using 
the \texttt{DISENTANGLING\_SHIFT\_AND\_ADD} code\footnote{\url{https://github.com/TomerShenar/Disentangling\_Shift\_And\_Add/}} 
\citep{Shenar+2020, Shenar+2022} over the wavelength range of 3800--5080 \AA, which includes the temperature indicators of 
H$_9$, H$_8$, \ion{Ca}{2} K, H$_{\rm \epsilon}$, H$_{\rm \delta}$, H$_{\rm \gamma}$, and H$_{\rm \beta}$. The initial parameters 
for the disentangling procedure, such as orbital ephemeris ($T_0$, $P_{\rm orb}$), eccentricity ($e$), and RV semi-amplitudes 
($K_1$, $K_2$), were obtained from the binary modeling in the following section. Unfortunately, the disentangled spectrum of 
the third component could not be extracted due to insufficient information about its orbital parameters. After running 
this procedure, the disentangled spectra of the primary and secondary stars were normalized using the \texttt{SUPPNET} package\footnote{\url{https://github.com/RozanskiT/suppnet/}} \citep{Rozanski+2022}. 


Using the \texttt{GSSP} software package\footnote{\url{https://fys.kuleuven.be/ster/meetings/binary-2015/gssp-software-package}} \citep{Tkachenko2015}, 
each disentangled spectrum was compared with synthetic spectra computed using the radiative transfer code \texttt{SYNTHV} \citep{Tsymbal1996} 
and a grid of \texttt{LLMODELs} stellar atmosphere models \citep{Shulyak+2004}. The TESS v8.2 \citep{Paegert+2022} and 
Gaia DR3 \citep{Gaia2023} catalogues reported the effective temperature ($T_{\rm eff}$) of TIC 322208686 as 7043$\pm$192 K and 
6731$_{-18}^{+16}$ K, respectively, so we used these as starting values. The surface gravities of the eclipsing components were 
set to $\log$ $g_{\rm A,B}$ = 4.2 from our modeling results. The micro- and macro-turbulence velocities were estimated by applying 
$T_{\rm eff}$, $\log$ $g$, and metallicities [M/H] to empirical relationships built in the \texttt{iSpec} code\footnote{\url{https://www.blancocuaresma.com/s/iSpec/}} 
\citep{Blanco-Cuaresma+2014}. 

\begin{figure}
\includegraphics[scale=0.9]{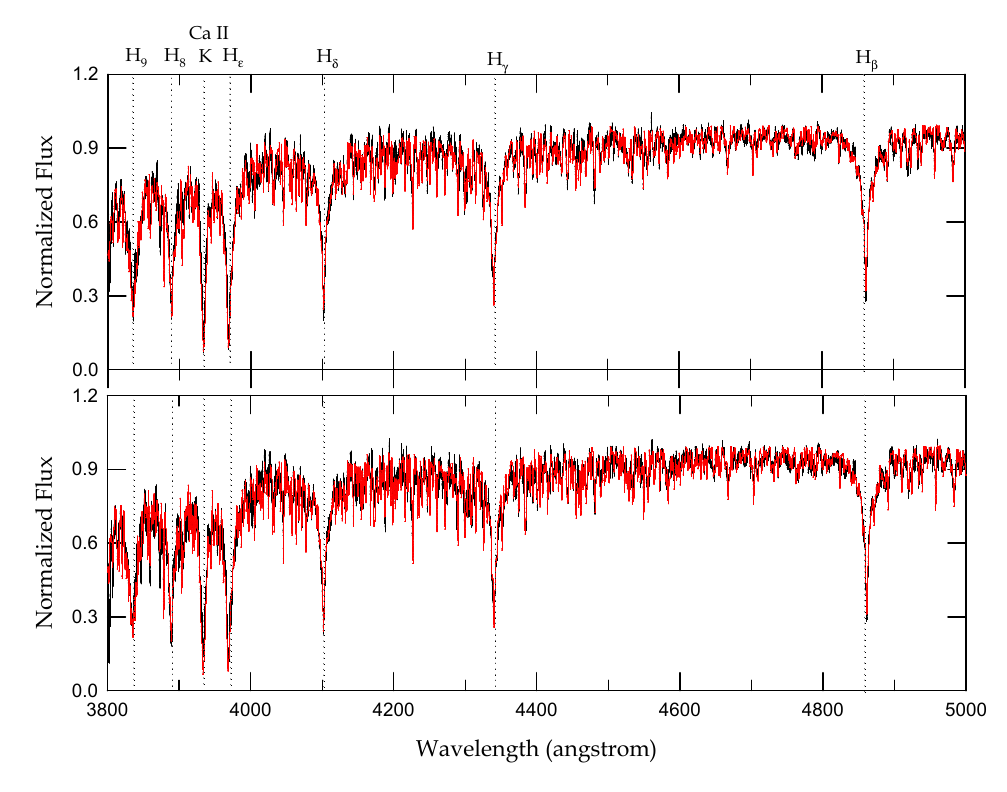}
\caption{Disentangled spectra of the primary (upper panel) and secondary (lower panel) components for TIC 322208686. The black 
and red solid lines represent the disentangled and best-fitting synthetic spectra, respectively. The vertical dotted lines 
represent the spectral lines labeled above the upper panel. }
\label{Fig5}
\end{figure}

In the \texttt{GSSP} runs, the effective temperature, rotational velocity, and metallicity were adjusted and evaluated. However, 
we could not determine the metallicities of both components, probably due to low SNR and the incompleteness of the disentangled spectra. 
Therefore, the analysis proceeded in two steps. In the first step, we used the solar metallicities for each star. 
In the second step, the metallicities were set to [M/H] = 0.3, as inferred from the comparisons with evolutionary models described 
in the last section. As a result of this run, the effective temperatures and rotational velocities were found to be optimal at 
$T_{\rm A,eff}$ = $7028\pm100$ K and $v_{\rm A}$sin$i$ = $76\pm6$ km s$^{-1}$ for the primary component, and at $T_{\rm B,eff}$ 
= $6904\pm210$ K and $v_{\rm B}$sin$i$ = $64\pm6$ km s$^{-1}$ for the secondary component, by fitting a polynomial to the lowest 
$\chi^2$ values \citep[cf.][]{Lehmann+2011}. The micro- and macro-turbulence velocities were obtained as 1.77 km s$^{-1}$ and 
14.65 km s$^{-1}$ for TIC 322208686 A, and 1.69 km s$^{-1}$ and 13.08 km s$^{-1}$ for its companion, respectively.
Figure \ref{Fig5} shows the disentangled and best-fit spectra of the primary and secondary components.

\section{Binary Modeling and Absolute Parameters} \label{sec_binarity}

Observing double-lined EBs provides a unique opportunity to directly and precisely measure their most fundamental parameters, 
such as masses and radii. Such measurements will help to constrain and better understand models of stellar structure and 
evolution. The TESS photometric and BOES spectroscopic observations reveal that TIC 322208686 is a triple-lined RV system 
exhibiting continuous and significant brightness variations due to multiple pulsations. In addition, the phase difference 
between Min I and Min II is around 0.50, and the $O-C$ diagram of the eclipse timings in Figure \ref{Fig2} shows that 
the two eclipse types are in phase with each other. These imply that the orbital eccentricity of the inner binary is nearly zero. 
The \texttt{Wilson-Devinney} (W-D) modeling code \citep{Wilson+1971, Kallrath2022} was adopted to compute our target star 
observables. Our RV measurements were modeled with the short-cadence TESS light curves (S24, S57, and S58) in the same way as 
for the hierarchical triple WASP 0346-21 \citep{Lee+2024}. 

For this modeling, the primary's temperature was fixed to $T_{\rm A,eff}$ = $7028\pm100$ K as measured from the BOES spectra, 
while the secondary's $T_{\rm B,eff}$ was adjusted to best fit the ultra-precise light curves. We set the rotation parameters 
of the eclipsing components to $F_{\rm A}$ = 1.33$\pm$0.11 and $F_{\rm B}$ = 1.18$\pm$0.11 from the ratio of the observed 
$v_{\rm obs} \approx v \sin$$i$ and the calculated synchronous rotation $v_{\rm sync}$. 
The limb-darkening parameters $x_{\rm A,B}$ and $y_{\rm A,B}$ for the logarithmic law were interpolated from the updated values 
of \citet{van1993} implemented in the W-D code. Since the public TESS photometry and our BOES spectroscopy were not performed 
simultaneously and the tertiary object was detected in the echelle spectra but its orbital parameters are unknown, these two 
types of datasets were solved separately and iteratively \citep{Lee+2024,Lee+2025}. 

\begin{figure}
\includegraphics[scale=0.9]{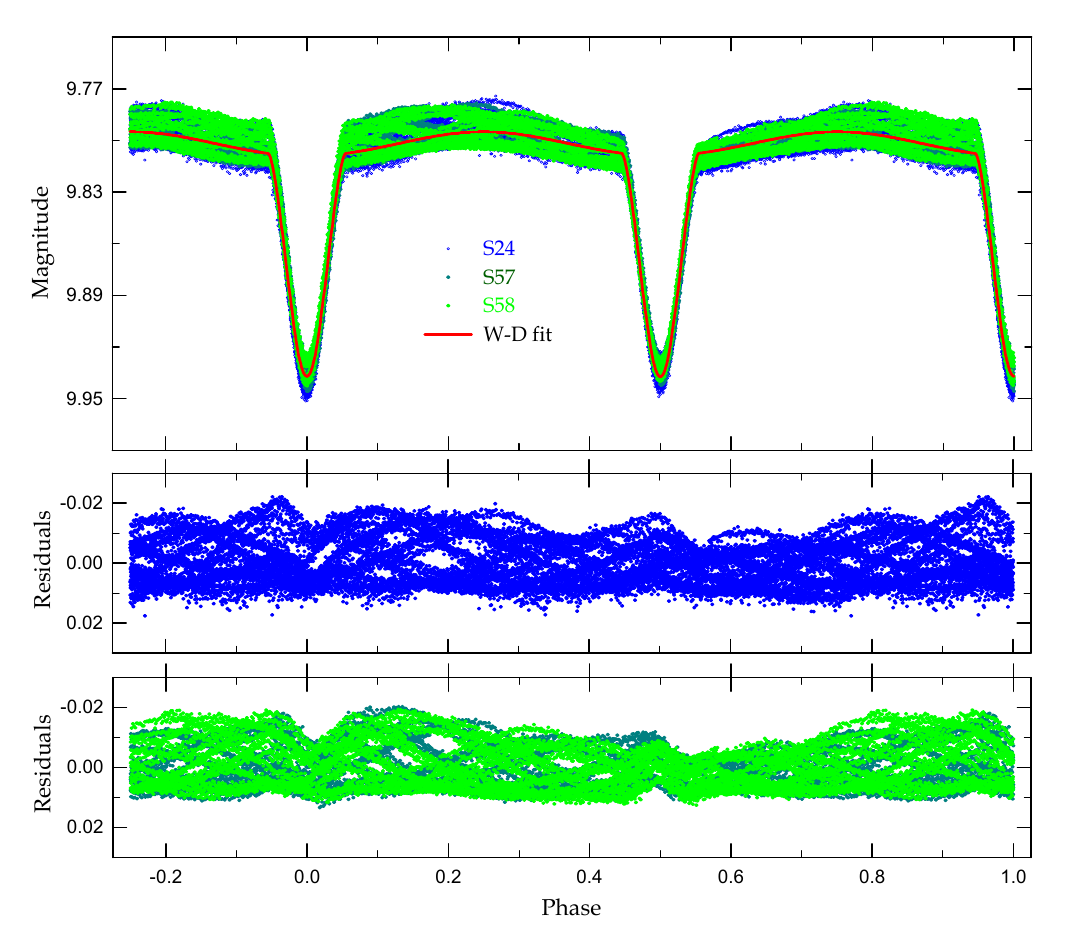}
\caption{The top panel displays the phased light curve of TIC 322208686 with the fitted model. The blue, cyan, and green circles 
are the TESS measurements from Sectors 24, 57, and 58, respectively, and the red solid curve is computed with our W-D fit. 
The middle and bottom panels represent the corresponding residuals for 120-s and 200-s cadence data, respectively, 
from the synthetic curve. }
\label{Fig6}
\end{figure}

The fitted parameters in the W-D run are: for the light curves, orbital ephemeris $T_0$ and $P_{\rm orb}$, inclination angle $i$, 
secondary's temperature $T_{\rm B,eff}$, dimensionless surface potentials $\Omega _{\rm A,B}$, and bandpass luminosities 
$l_{\rm A}$ and $l_3$; and for the RV curves, reference epoch $T_0$, orbital semi-major axis $a$, systemic velocity $\gamma$, and 
mass ratio $q$. The resulting parameters from the binary modeling are presented in Table \ref{Tab3}, and the corresponding model fits 
are superimposed on the observed light and RV curves as gray and red solid lines in Figures \ref{Fig1} and \ref{Fig4}, respectively. 
In addition, the phase-folded light curve and the W-D model residuals are shown in Figure \ref{Fig6}. 
The agreement between the binary model and the observations is very good. To yield the parameter errors in Table \ref{Tab3}, 
we applied numerous models with different inputs of the initial parameters and various approaches. From these runs, we estimated 
the errors using the variation in each fitted parameter \citep{Southworth+2022a}.  

As presented in the next section, our target star exhibited multiperiod pulsations, so we investigated whether they affected 
the binary parameters. First, we found the pulsating frequencies from all residual lights, including the data of both eclipse 
phases. We then analyzed the pre-whitened EB light curve, removing these pulsations, using the W-D code. The results are in 
good agreement with those from observed TESS data within their errors. 
The absolute properties of TIC 322208686 were computed using the parameters in Table \ref{Tab3} obtained from the joint light and RV fits.
We assigned the solar temperature and bolometric magnitude as $T_{\rm eff}$$_\odot$ = 5780 K and $M_{\rm bol}$$_\odot$ = +4.73, 
respectively. The calculation results are summarized in Table \ref{Tab4}, where the bolometric corrections (BC) were adopted from 
empirical relationships with the effective temperatures of the component stars \citep{Flower1996,Torres2010}. 

The distance to an EB system is usually calculated via the distance-modulus equation $V-A_{\rm V}-M_{\rm V}=5\log{d}-5$, 
with the apparent magnitude $V$ taken at maximum light and the interstellar extinction $A_{\rm V}$ $\simeq$ 3.1$E(B-V)$. 
For our target TIC 322208686, the TESS v8.2 catalogue \citep{Paegert+2022} gave $V$ = +9.685$\pm$0.004, ($B-V$) = 
+0.372$\pm$0.051, and $E(B-V)$ = 0.054$\pm$0.028. The TESS $V$ magnitude would be the combined brightness of the triple system, 
including the third companion detected in this study. Assuming that the third-light contribution derived from our TESS modelling and 
the BFs approximately reflects the $V$-band contribution, we calculated an apparent magnitude of $V_{\rm EB}$ = 9.870 
for the eclipsing pair of TIC 322208686. Using $V_{\rm EB}$, the TESS color index, and our modeling parameters in Tables \ref{Tab3} 
and \ref{Tab4}, we estimated a geometric distance of $d$ = 323$\pm$13 pc for TIC 322208686 AB, which is 53$\pm$14 pc closer than 
376$\pm$6 pc inverted from the Gaia DR3 parallax of $\pi$ = 2.661$\pm$0.041 mas \citep{Gaia2023}. 
Ignoring interstellar reddening ($A_{\rm V}$ = 0.0), the EB distance increases to 349$\pm$14pc, but is still smaller than the GAIA 
distance. The discrepancy between the EB and GAIA distances may be partly attributable to brightness variations caused by pulsations. 

The Gaia parameter RUWE indicates the goodness-of-fit of the astrometric solution for this spacecraft and is quite sensitive 
to unresolved close companions \citep{Stassun+2021}. TIC 322208686 has a RUWE value of 3.181, which is considerably larger 
than the good solution of $\la 1.0$. RUWE values higher than 1.4 indicate that the astrometric fits are unreliable and do not 
perfectly describe the motion of the objects. This may be caused by the presence of the tight tertiaries in binary systems. 
The unusual RUWE value in TIC 322208686 may be due to a third companion star detected in our BOES spectrum.

\section{Pulsational Characteristics} \label{sec_pulsation}

\begin{figure}
\includegraphics[scale=0.9]{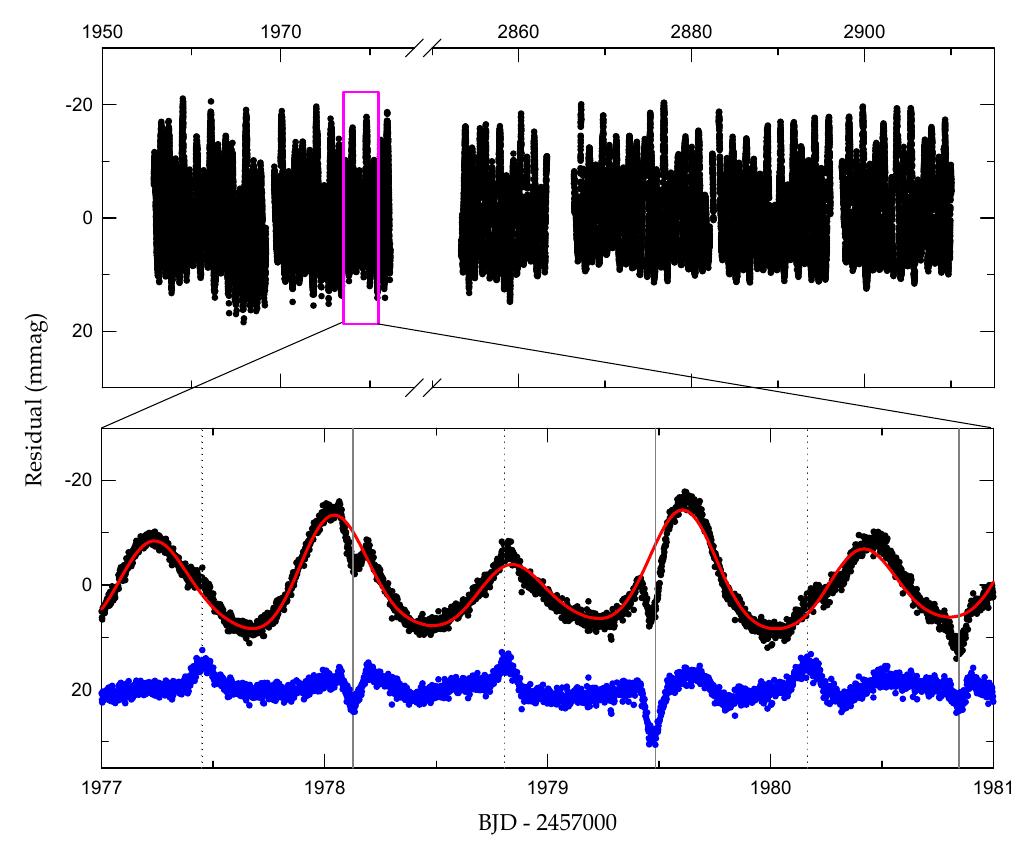}
\caption{Light curve residuals from the W-D model fit to the TESS data. The lower panel is a zoomed-in view of the residuals 
shown in the inset box in the upper panel. The red solid line represents the synthetic curve computed from the 11-frequency fit 
to the out-of-eclipse data. The residuals from the frequency fit are offset from zero by +20 mmag for clarity and are plotted as 
the blue symbols. The vertical solid and dotted lines indicate the primary and secondary minimum epochs, respectively. }
\label{Fig7}
\end{figure}

\begin{figure}
\includegraphics[]{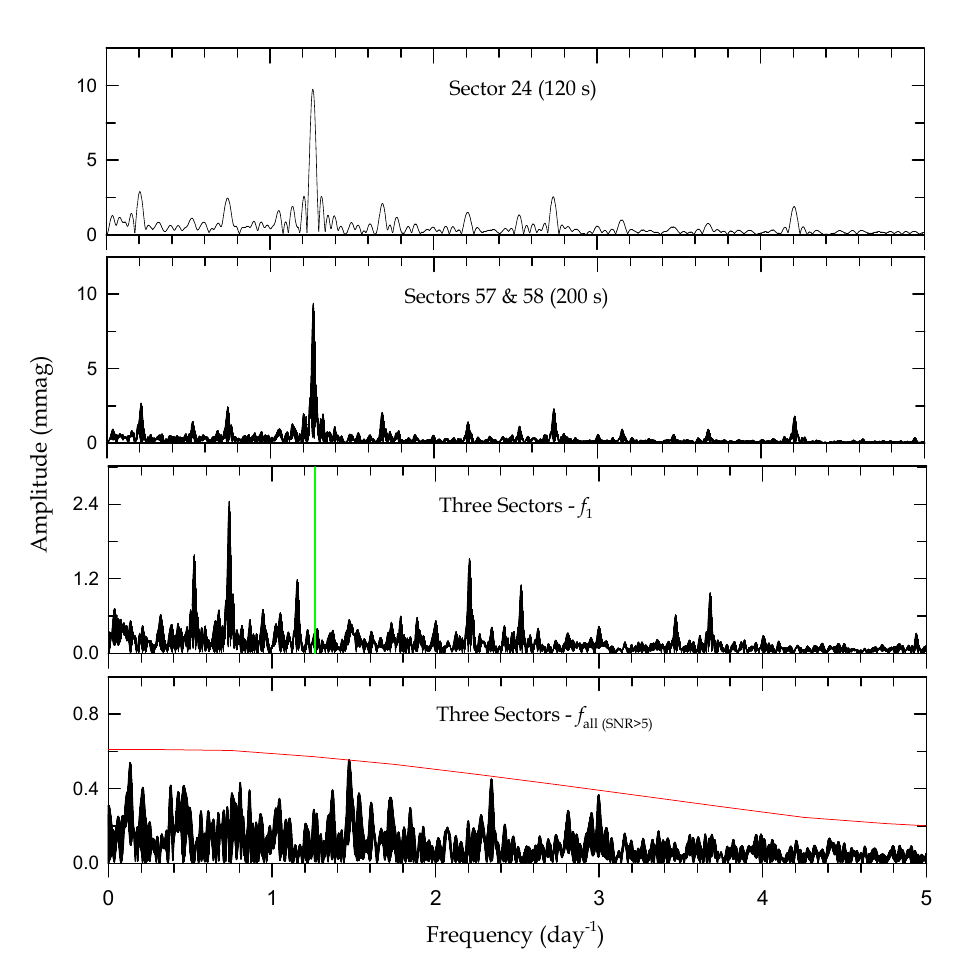}
\caption{\texttt{PERIOD04} periodograms for the outside-eclipse TESS residuals. The first and second panels show the amplitude spectra 
of the 120-s and 200-s residuals, respectively, while the third and bottom panels are obtained using the entire short-cadence data, 
pre-whitening the most dominant frequency ($f_1$) as indicated by a vertical green line and all 11 frequencies with SNR $> 5$, 
respectively. The red solid line in the bottom panel corresponds to five times the noise spectrum. }
\label{Fig8}
\end{figure}

The fundamental parameters in Table \ref{Tab4} indicate that the eclipsing components are main-sequence dwarfs with spectral type 
$\sim$F1 V and are therefore candidates for intermediate-mass pulsators such as $\gamma$ Dor or $\delta$ Sct variables \citep{Antoci+2019}. 
Figure \ref{Fig7} presents the TESS light residuals of the W-D model fit, with the lower panel providing more details on the residuals. 
This figure shows distinct brightness variations with a main period of about 0.80 days and a maximum amplitude of $>20$ mmag. 
To reliably extract the pulsation signals, we used the out-of-eclipse residuals with the binary effects removed. 
The frequency search was conducted using the \texttt{PERIOD04} computer code \citep{Lenz+2005} up to the Nyquist frequency 
$f_{\rm Ny}$ = 216 day$^{-1}$, which corresponds to a sampling rate of 200 s for S57 and S58. We first analyzed each of two 
datasets, S24 and S57$-$58, separated by a long observation gap of more than 2 yrs. The resultant amplitude spectra are presented 
in the first and second panels of Figure \ref{Fig8}. There are no noticeable difference in amplitudes or frequencies between them.

Our frequency analysis was finally applied to the full short-cadence residuals, excluding the two eclipse periods, to find 
consistent pulsation signals. As a consequence, we extracted 11 significant frequencies with SNR greater than 5 
\citep{Baran+2021,Lee+2025} using an iterative pre-whitening method. In this process, the residual light curves 
were pre-whitened to the next frequency after finding the frequency \citep{Lee+2014}.  
The third and bottom panels of Figure \ref{Fig8} display the Fourier amplitude spectra for the three sectors' residuals, respectively, 
with the most dominant $f_1$ frequency and all 11 frequencies pre-whitened. We did not find any significant signals in 
the frequency range of $> 5$ day$^{-1}$, which is not displayed in this figure.

The multi-frequency parameters for TIC 322208686 are summarized in Table \ref{Tab5}, and the resulting synthetic curve is plotted 
as a red solid line in the lower panel of Figure \ref{Fig7}. The residual light curves from the 11-frequency fit, 
with both binary and pulsation effects removed, are offset by +20 mmag and appear as blue dots on the same panel. This figure 
reveals that the in-eclipse residuals deviate from the general trend of the outside-eclipse residuals. During the primary eclipses, 
the residual light curves get darker until conjunctions and then brighten again, whereas during the secondary eclipses, 
the reverse occurs: they brighten to conjunctions and then darken. These features suggest that the light contribution from pulsations 
to the EB system is relatively largest at the secondary minimum among the entire binary orbit, and thus the primary star maximally obscuring 
the secondary companion at Min II is a pulsating variable. 

In order to uncover aliasing signals such as frequency harmonic and combination terms, the assigned frequency resolution $F_{\rm res}$ 
was 1.5/$\Delta T$ = 0.00157 day$^{-1}$ using the duration $\Delta T$ = 954.3 days of the TESS data used \citep{Loumos+1978}. 
The search results are presented on the remark column in Table \ref{Tab5}. Within the $F_{\rm res}$ criterion, frequencies $f_3$, 
$f_6$, and $f_9$ are combinations with the detected signals and/or orbital frequency ($f_{\rm orb}$ = 0.73566 day$^{-1}$), while 
$f_5$ is twice the most prominent frequency $f_1$. Also, the lowest frequencies $f_7$ and $f_{10}$ appear to be artifacts related 
to the 13.7-day TESS orbit. The remaining frequencies can be considered independent signals, and their pulsation constants 
were obtained from the relationship $Q_{\rm i}$ = $f_{\rm i}$$\sqrt{\rho_{\rm A} / \rho_\odot}$, as follows: $Q_{1}$ = 0.496 days, 
$Q_{2}$ = 0.847 days, $Q_{4}$ = 0.542 days, $Q_{8}$ = 0.951 days, and $Q_{11}$ = 0.332 days. The pulsation frequencies and 
$Q$ values are within the typical ranges of $\gamma$ Dor variables \citep{Grigahcene+2010,Uytterhoeven+2011}. The frequency ratio 
of $f_{\rm orb}$:$f_1$ $\simeq$ 3:5 suggests that the dominant $\gamma$ Dor pulsation may arise from tidal interactions between 
the EB components.

\section{Discussion and Conclusion} \label{sec_conclusion}

In this study, we present the first-measured spectroscopic characterizations of TIC 322208686 with high-quality TESS data. 
The BOES echelle spectra showed the presence of not only the eclipsing pair but also a tertiary companion. We obtained 
the RV measurements of the three objects using the BF profiles with rotation functions applied. While the eclipsing components 
moved with similar RV semi-amplitudes ($K_{\rm A,B}$) in a 1.3593-day circular orbit, the tertiary object showed little change 
during the observation run, with an average velocity of $-13.86\pm0.37$ km s$^{-1}$. The disentangled spectra for 
each EB component were extracted using the \texttt{DISENTANGLING\_SHIFT\_AND\_ADD} code, and compared to synthetic model grids 
to determine the effective temperatures and rotational rates for solar metallicity and [M/H] = 0.3. The spectral analyses revealed 
that TIC 322208686 is a triple-lined star system consisting of an inner F1 V pair and a distant tertiary. 

We synthesized the TESS light and our RV curves of the triple system, including the atmospheric parameters, to characterize 
its physical properties of $M_{\rm A} = 1.564\pm0.012$ $M_\odot$, $R_{\rm A} = 1.588\pm0.011$ $R_\odot$, $T_{\rm A,eff}$ = 
$7028\pm100$ K, and $L_{\rm A} = 5.51\pm0.32$ $L_\odot$ for TIC 322208686 A and $M_{\rm B} = 1.483\pm0.012$ $M_\odot$, $R_{\rm B} 
= 1.500\pm0.012$ $R_\odot$, $T_{\rm B,eff}$ = $7020\pm110$ K, and $L_{\rm B} = 4.90\pm0.32$ $L_\odot$ for TIC 322208686 B, 
respectively. The binary modeling allowed us to derive the masses and radii of the eclipsing pair to less than 1 \% accuracy. 
Our measurements of $v_{\rm A,B}\sin$$i$ indicate that the eclipsing components A and B are expected to be in a state of 
supersynchronization at present. The light fractions of the three components, calculated via this synthesis, agree well with 
the average values obtained from the BOES spectra.

\begin{figure}
\includegraphics[scale=0.9]{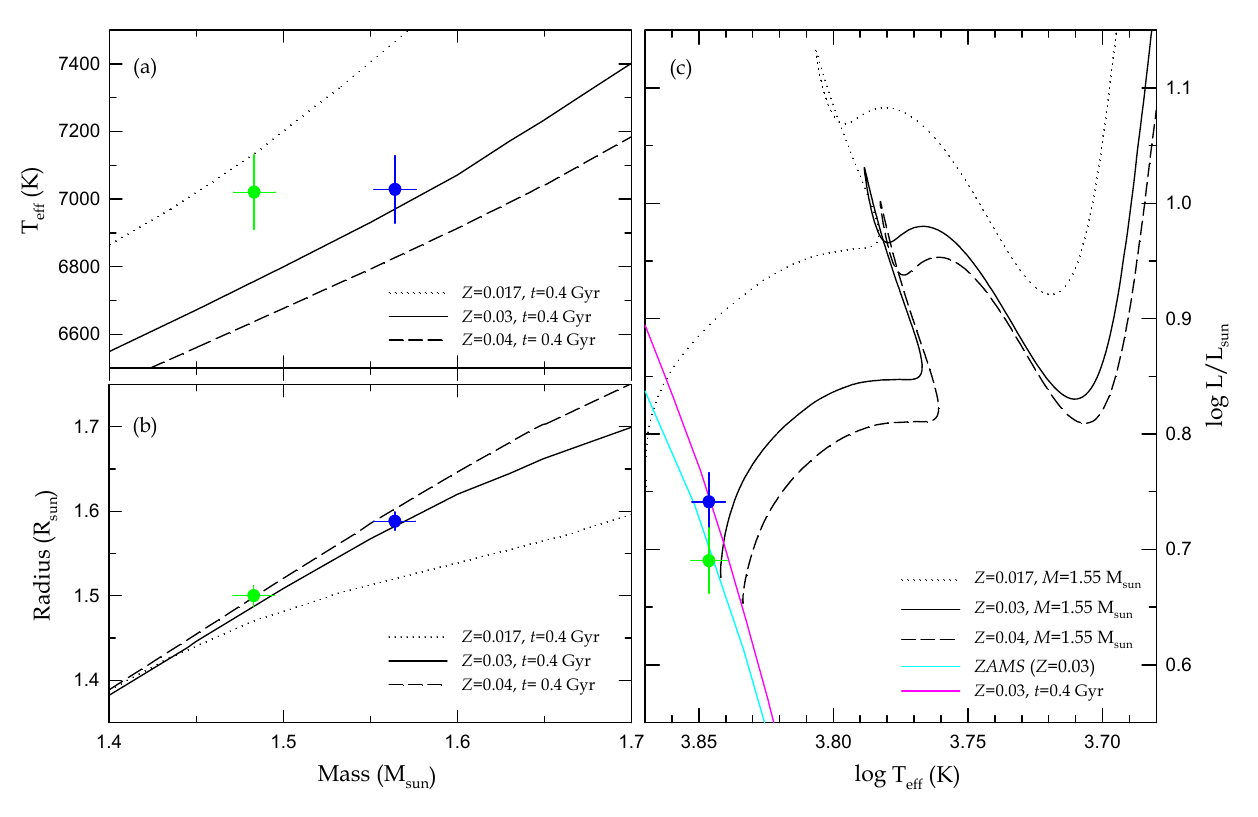}
\caption{Position of TIC 322208686 A (blue circle) and B (green circle) on the mass-temperature (a), mass-radius (b), and 
H-R (c) diagrams. In all panels, the dotted, solid, and dashed lines are the theoretic predictions with different metallicities 
of $Z$ = 0.017, 0.03, and 0.04, respectively, from the PARSEC models \citep{Bressan+2012}. In the right panel, the ZAMS and 
0.4-Gyr isochrone for $Z$ = 0.03 are presented as cyan and red solid lines, respectively. }
\label{Fig9}
\end{figure}

Comparisons of the EB fundamental parameters with stellar models provide a straightforward and clear interpretation of 
the agreement between observations and theory. In particular, since mass, radius, and temperature are physical properties that 
can be measured directly through observations, one of the most effective methods would be to compare them to the model predictions. 
To study the evolutionary status of TIC 322208686, we first plotted the positions of the EB components on the mass-radius ($M-R$) 
and mass-temperature ($M-T_{\rm eff}$) diagrams in the left panels of Figure \ref{Fig9}. Here, the dotted, solid, and dashed lines 
are the PARSEC v$1.2S$ stellar evolution models\footnote{\url{https://stev.oapd.inaf.it/PARSEC/}} \citep{Bressan+2012} with 
metal abundances $Z$ of 0.017 (solar value), 0.03, and 0.04, respectively. We found an acceptable fit for a model with an age of 
$t$ = 0.4 Gyr and a metallicity of $Z$ = 0.03, which is more metal-rich than the solar composition used in our initial spectral analysis. 
Our choice of age and metallicity is further supported by the $T_{\rm eff}-L$ plot in the right panel of Figure \ref{Fig9}. 
The cyan and red solid lines denote the zero-age main sequence (ZAMS) and the 0.4-Gyr isochrone, both with $Z$ = 0.03. In the H-R diagram, 
the A component is slightly evolved from ZAMS, while the B companion, which is about 0.08 $M_\odot$ lighter, is located on ZAMS. 


For Galactic kinematics and population membership, we calculated the space velocities of TIC 322208686 as $U = 27.8\pm0.7$ km s$^{-1}$, 
$V = 235.6\pm0.7$ km s$^{-1}$, and $W = -1.1\pm0.3$ km s$^{-1}$, using the same approach as for \citet{Lee+2024} based on 
the formulation of \citet{Johnson+1987}. In this calculation, we used the Gaia DR3 measurements and our $\gamma$ velocity and 
distance. The EB position in the $U-V$ diagram suggests that the target star has the kinematics of the thin-disk population 
\citep{Pauli+2006}. The classification of TIC 322208686 as a thin-disk star is consistent with the metallicity of $Z$ = 0.03 
predicted by stellar evolution models.

\begin{acknowledgements}
This paper is based on the echelle spectra obtained from BOAO and the light curves made by the TESS mission. 
The short-cadence TESS data were collected through Guest Investigator Program G022208 (PI: Nadia Zakamska). 
We wish to thank the valuable comments and suggestions of the anonymous referee. We gratefully acknowledge the support by 
the KASI grant 2025-1-830-05. H.-Y.K. was supported by the grant number RS-2023-00271900 from the National Research Foundation 
(NRF) of Korea. 
\end{acknowledgements}

\clearpage 
\begin{deluxetable}{lcccc}
\tablewidth{0pt}
\tablecaption{TESS Eclipse Timings of TIC 322208686 \label{Tab1}}
\tablehead{
\colhead{BJD}    & \colhead{Error} & \colhead{Epoch} & \colhead{$O-C$} & \colhead{Min}  
}
\startdata
2,458,739.57378  & $\pm$0.00021    & $-159.5$        & $+0.00230$      & II            \\
2,458,740.25376  & $\pm$0.00036    & $-159.0$        & $+0.00263$      & I             \\
2,458,740.93124  & $\pm$0.00015    & $-158.5$        & $+0.00045$      & II            \\
2,458,741.61026  & $\pm$0.00068    & $-158.0$        & $-0.00019$      & I             \\
2,458,742.29154  & $\pm$0.00044    & $-157.5$        & $+0.00144$      & II            \\
2,458,742.97190  & $\pm$0.00025    & $-157.0$        & $+0.00214$      & I             \\
2,458,743.65167  & $\pm$0.00056    & $-156.5$        & $+0.00225$      & II            \\
2,458,745.01041  & $\pm$0.00038    & $-155.5$        & $+0.00168$      & II            \\
2,458,745.68781  & $\pm$0.00079    & $-155.0$        & $-0.00058$      & I             \\
2,458,746.36872  & $\pm$0.00015    & $-154.5$        & $+0.00067$      & II            \\
\enddata
\tablecomments{This table is available in its entirety in machine-readable form. A portion is shown here for guidance regarding its form and content.}
\end{deluxetable}

\begin{deluxetable}{lccccccc} 
\tabletypesize{\small}
\tablewidth{0pt}                    
\tablecaption{Radial velocities of TIC 322208686$\dag$ \label{Tab2}}                                                                            
\tablehead{ 
\colhead{BJD}          & \colhead{$V_{\rm A}$}   & \colhead{$\sigma_{\rm A}$} & \colhead{$V_{\rm B}$}   & \colhead{$\sigma_{\rm B}$} & \colhead{$V_{\rm C}$}   & \colhead{$\sigma_{\rm C}$}    \\ 
\colhead{(2,460,000+)} & \colhead{(km s$^{-1}$)} & \colhead{(km s$^{-1}$)}    & \colhead{(km s$^{-1}$)} & \colhead{(km s$^{-1}$)}    & \colhead{(km s$^{-1}$)} & \colhead{(km s$^{-1}$)}    
}
\startdata                                                                                                                       
260.9055               & $+120.80$                & 0.40                       & $-131.29$               & 0.79                       & $-14.35$                & 0.52                          \\ 
260.9267               & $+124.80$                & 0.40                       & $-135.57$               & 0.80                       & $-13.29$                & 0.53                          \\ 
260.9479               & $+127.89$                & 0.37                       & $-138.38$               & 0.79                       & $-14.58$                & 0.53                          \\ 
260.9690               & $+129.00$                & 0.44                       & $-140.30$               & 0.84                       & $-14.19$                & 0.53                          \\ 
260.9902               & $+129.47$                & 0.44                       & $-139.84$               & 0.82                       & $-13.96$                & 0.50                          \\ 
261.0114               & $+127.92$                & 0.46                       & $-139.18$               & 0.82                       & $-13.95$                & 0.58                          \\ 
261.0326               & $+126.86$                & 0.47                       & $-136.75$               & 0.85                       & $-14.09$                & 0.56                          \\ 
262.9139               & $-113.47$                & 0.40                       & $+117.50$               & 0.95                       & $-13.32$                & 0.54                          \\ 
262.9350               & $-118.92$                & 0.38                       & $+121.26$               & 0.94                       & $-13.97$                & 0.51                          \\ 
262.9562               & $-125.16$                & 0.38                       & $+127.99$               & 0.83                       & $-13.56$                & 0.58                          \\ 
262.9774               & $-128.38$                & 0.37                       & $+132.33$               & 0.83                       & $-13.65$                & 0.58                          \\ 
262.9986               & $-130.70$                & 0.36                       & $+134.27$               & 0.80                       & $-13.38$                & 0.53                          \\ 
263.0198               & $-133.00$                & 0.34                       & $+136.74$               & 0.81                       & $-13.76$                & 0.53                          \\ 
263.0409               & $-132.67$                & 0.34                       & $+135.36$               & 0.84                       & $-13.53$                & 0.52                          \\ 
263.0621               & $-131.38$                & 0.34                       & $+134.68$               & 0.86                       & $-14.21$                & 0.51                          \\ 
263.0833               & $-128.84$                & 0.34                       & $+131.30$               & 0.77                       & $-13.97$                & 0.52                          \\ 
\enddata 
\tablenotetext{\dag}{$V_{\rm A}$, $V_{\rm B}$, and $V_{\rm C}$ represent the measured RVs of the primary, secondary, and third stars, respectively, and $\sigma_{\rm A}$, $\sigma_{\rm B}$, and $\sigma_{\rm C}$ are their uncertainties. }
\end{deluxetable}

\begin{deluxetable}{lcc}
\tablewidth{0pt} 
\tablecaption{Light and RV Parameters of TIC 322208686 \label{Tab3}}
\tablehead{
\colhead{Parameter}               & \colhead{Primary (A)} & \colhead{Secondary (B)}                                                  
}                                                                                     
\startdata 
$T_0$ (BJD)                       & \multicolumn{2}{c}{2,458,956.37822$\pm$0.00030}     \\
$P_{\rm orb}$ (day)               & \multicolumn{2}{c}{1.35931936$\pm$0.00000060}       \\
$i$ (deg)                         & \multicolumn{2}{c}{75.40$\pm$0.16}                  \\
$T_{\rm eff}$ (K)                 & 7028$\pm$100          & 7020$\pm$110                \\
$\Omega$                          & 5.709$\pm$0.026       & 5.790$\pm$0.029             \\
$\Omega_{\rm in}$$\rm ^a$         & \multicolumn{2}{c}{3.666}                           \\
$F$                               & 1.33$\pm$0.11         & 1.18$\pm$0.11               \\
$X$, $Y$                          & 0.640, 0.254          & 0.640, 0.253                \\
$x$, $y$                          & 0.527, 0.282          & 0.527, 0.282                \\
$l/(l_{\rm A}+l_{\rm B}+l_3)$     & 0.4461$\pm$0.0049     & 0.3967                      \\
$l_3$$\rm ^b$                     & \multicolumn{2}{c}{0.1572$\pm$0.0039}               \\
$r$ (pole)                        & 0.2092$\pm$0.0013     & 0.1982$\pm$0.0015           \\
$r$ (point)                       & 0.2163$\pm$0.0015     & 0.2037$\pm$0.0017           \\
$r$ (side)                        & 0.2126$\pm$0.0014     & 0.2005$\pm$0.0016           \\
$r$ (back)                        & 0.2153$\pm$0.0015     & 0.2029$\pm$0.0017           \\
$r$ (volume)$\rm ^c$              & 0.2124$\pm$0.0014     & 0.2006$\pm$0.0016           \\ [1.0mm]
\multicolumn{3}{l}{Spectroscopic orbits:}                                               \\
$T_0$ (BJD)                       & \multicolumn{2}{c}{2,460,261.3299$\pm$0.0011}       \\
$a$ (R$_\odot$)                   & \multicolumn{2}{c}{7.483$\pm$0.019}                 \\
$\gamma$ (km s$^{-1}$)            & \multicolumn{2}{c}{$-1.84\pm$0.73}                  \\
$K$ (km s$^{-1}$)                 & 131.24$\pm$0.45       & 138.39$\pm$0.46             \\
$q$                               & \multicolumn{2}{c}{0.9483$\pm$0.0045}               \\
\enddata                                                                                
\tablenotetext{a}{Potential for the inner critical Roche surface.}
\tablenotetext{b}{Value at 0.25 orbital phase. }
\tablenotetext{c}{Mean volume radius.}
\end{deluxetable}

\begin{deluxetable}{lcc}
\tablewidth{0pt} 
\tablecaption{Absolute Parameters of TIC 322208686 \label{Tab4}}
\tablehead{
\colhead{Parameter}           & \colhead{Primary (A)} & \colhead{Secondary (B)}                                                  
}                                                                                                                                     
\startdata                                                                                                                            
$M$ ($M_\odot$)               & 1.564$\pm$0.012       & 1.483$\pm$0.012           \\
$R$ ($R_\odot$)               & 1.588$\pm$0.011       & 1.500$\pm$0.012           \\
$\log$ $g$ (cgs)              & 4.230$\pm$0.007       & 4.257$\pm$0.008           \\
$\rho$ ($\rho_\odot$)         & 0.391$\pm$0.009       & 0.440$\pm$0.011           \\
$v_{\rm sync}$ (km s$^{-1}$)  & 59.11$\pm$0.42        & 55.86$\pm$0.47            \\
$v$$\sin$$i$ (km s$^{-1}$)    & 76$\pm$6              & 64$\pm$6                  \\
$T_{\rm eff}$ (K)             & 7028$\pm$100          & 7020$\pm$110              \\
$L$ ($L_\odot$)               & 5.51$\pm$0.32         & 4.90$\pm$0.32             \\
$M_{\rm bol}$ (mag)           & 2.87$\pm$0.06         & 3.00$\pm$0.08             \\
BC (mag)                      & 0.03$\pm$0.01         & 0.03$\pm$0.01             \\
$M_{\rm V}$ (mag)             & 2.84$\pm$0.06         & 2.97$\pm$0.08             \\
$E(B-V)$ (mag)                & \multicolumn{2}{c}{0.054$\pm$0.028}               \\
Distance (pc)                 & \multicolumn{2}{c}{323$\pm$13}                    \\
\enddata
\end{deluxetable}

\begin{deluxetable}{lrccrc}
\tablewidth{0pt}
\tablecaption{Multi-frequency Solution for TIC 322208686$\rm ^{a,b}$ \label{Tab5}}
\tablehead{
             & \colhead{Frequency}    & \colhead{Amplitude} & \colhead{Phase} & \colhead{SNR$\rm ^c$} & \colhead{Remark}    \\
             & \colhead{(day$^{-1}$)} & \colhead{(mmag)}    & \colhead{(rad)} &                       &             
} 
\startdata 
$f_{1}$      &  1.26168$\pm$0.00001   & 9.44$\pm$0.20       & 2.83$\pm$0.06   & 82.65                 &                     \\
$f_{2}$      &  0.73787$\pm$0.00001   & 3.17$\pm$0.21       & 2.95$\pm$0.19   & 26.20                 &                     \\
$f_{3}$      &  0.52490$\pm$0.00002   & 1.86$\pm$0.21       & 3.84$\pm$0.33   & 15.24                 & $f_1-f_{\rm orb}$   \\
$f_{4}$      &  1.15402$\pm$0.00003   & 1.26$\pm$0.20       & 4.52$\pm$0.46   & 10.88                 &                     \\
$f_{5}$      &  2.52335$\pm$0.00003   & 1.10$\pm$0.15       & 4.94$\pm$0.41   & 12.26                 & 2$f_1$              \\
$f_{6}$      &  1.99951$\pm$0.00004   & 0.82$\pm$0.17       & 1.95$\pm$0.62   &  8.09                 & $f_5-f_3$           \\
$f_{7}$      &  0.03486$\pm$0.00006   & 0.75$\pm$0.21       & 5.80$\pm$0.82   &  6.10                 & artifact            \\
$f_{8}$      &  0.65720$\pm$0.00006   & 0.72$\pm$0.21       & 5.52$\pm$0.84   &  5.94                 &                     \\
$f_{9}$      &  1.78653$\pm$0.00005   & 0.74$\pm$0.18       & 3.93$\pm$0.72   &  7.01                 & $f_1+f_3$           \\
$f_{10}$     &  0.07581$\pm$0.00007   & 0.61$\pm$0.21       & 3.63$\pm$1.00   &  5.01                 & artifact            \\
$f_{11}$     &  1.88617$\pm$0.00006   & 0.58$\pm$0.18       & 5.96$\pm$0.90   &  5.59                 &                     \\
\enddata
\tablenotetext{a}{Frequencies are listed in order of detection. }
\tablenotetext{b}{Parameters' errors were obtained following Kallinger et al. (2008). }
\tablenotetext{c}{Calculated in a range of 5 day$^{-1}$ around each frequency. }
\end{deluxetable}

\end{document}